\documentclass[aps,showpacs,preprint]{revtex4}
\usepackage{amstext}
\usepackage{amssymb}
\usepackage{graphicx}

\begin{document}

\title{$A$ dependence of the enhancement factor in energy-weighted sums for
isovector giant resonances}
\author{V. M. Kolomietz and S. V. Lukyanov}
\affiliation{Institute for Nuclear Research, 03680 Kiev, Ukraine}

\begin{abstract}
We consider the energy weighted sums (EWS) for isovector giant dipole
resonances (IVGDR) in finite nuclei within Landau kinetic theory. The
dependence of both IVGDR energy, $E_{\mathrm{IVGDR}}$, and the EWS enhancement
factor, $\kappa (A)$, on the mass number $A$ occurs because of the boundary
condition on the moving nuclear surface. The values of $E_{\mathrm{IVGDR}} 
A^{1/3}$ and $\kappa (A)$ increase with $A$. The obtained value of the
enhancement factor is about $10\%$ for light nuclei and reaches
approximately $20\%$ for heavy nuclei. A fit of the enhancement factor to
the proper experimental data provides a value for the isovector Landau
amplitude of $F_{1}^{\prime }\simeq 1.1$.
\end{abstract}

\pacs{21.60.Ev, 24.30.Cz}
\maketitle

\section{Introduction}

A microscopic description of the isovector excitations within the nuclear
Fermi-liquid theory requires the use of two Landau amplitudes $F_{0}^{\prime
}$ and $F_{1}^{\prime }$ to model the nucleon-nucleon interaction. The
amplitude $F_{0}^{\prime }$ provides the isospin symmetry energy, whereas the
inclusion of the velocity-dependent force $\sim F_{1}^{\prime }$ leads to
the renormalization of the isovector energy weighted sums (EWS) $m_{k}$ \cite%
{LiSt,HaSaZh}. In particular, the isovector EWS $m_{1}$ becomes dependent on
the nucleon-nucleon interaction. This is opposite to the case of the
isoscalar excitations where the corresponding sum $m_{1}$ is model
independent. Moreover, the presence of the velocity dependent force gives
rise to an exceed of the $100\%$ exhaustion of Thomas-Reiche-Khun (TRK) sum
rule for the isovector giant dipole resonances (IVGDR). The origin of the
corresponding enhancement factor of the IVGDR sum $m_{1}$ was intensively
investigated for both the nuclear matter and the finite nuclei, see Refs. 
\cite{LiSt,tror87} and references therein. The RPA calculations of the
enhancement factor for the symmetric nuclear matter were recently performed
for a representative set of Skyrme forces in Ref. \cite{nest08}. As shown in 
\cite{nest08}, the value of the enhancement factor is changed in almost 2
times depending on the choice of the Skyrme force parametrization. The high
sensitivity of the enhancement factor to the choice of the Skyrme forces was
also demonstrated in Ref. \cite{trco08} for the nucleus $^{208}$Pb.

In finite nuclei, both the IVGDR eigenenergy $E_{\mathrm{IVGDR}}$ and the
EWS $m_{1}$ are rather complicated functions of the mass number $A$. In
contrast to the classical Steinwedel-Jensen model \cite{BoMo}, the value $E_{%
\mathrm{IVGDR}}\cdot A^{1/3}$ is not a constant but increases with $A$. The
theoretical approaches to microscopic description of IVGDR are mainly based
on the Random Phase Approximation (RPA) \cite{RiSh}. The RPA analysis of the
enhancement factor for some spherical nuclei (but not its $A$-dependence)
has been recently done in Ref. \cite{sabo04}. Note, however, that the RPA
calculations \cite{LiSt,tror87,sabo04,krtr80} overestimate the magnitude of
the enhancement factor, see below \figurename\ \ref{fig3}. Note also, that
within the RPA, the highly excited IVGDR is strongly fragmented over a wide
energy interval and a special averaging procedure has to be applied to pick
up a smooth $A$-dependence of the IVGDR characteristics. Furthermore, the
RPA calculations of the EWS $m_{k}$ are restricted due to taking into
account $1p-1h$ excitations only and, thereby, one can expect an
underestimation of the contribution to $m_{1}$ from more complicated states 
\cite{LiSt,befl75,krtr80,lilu91}.

In this work, we study the IVGDR within the Landau kinetic theory \cite%
{AbKha}, that is extended to the finite two-component Fermi-liquid drop. The 
$A$-dependence of \ both the IVGDR eigenenergy and the corresponding EWS
occurs due to the boundary conditions on the moving nuclear surface. Our
approach is more general than the scaling model \cite{stri83} or the fluid
dynamic approaches \cite{hoec79} due to the fact that it takes into
consideration all multipolarities of the Fermi surface distortions.

In Sect. II, we start from the response theory based on the collisionless
kinetic Landau-Vlasov approach to the isovector excitations. We derive the
main characteristics of the IVGDR, taking into account the velocity
dependent part of the Landau isovector interaction. In Sect. III, we obtain
the boundary condition for the isovector sound mode on the free nuclear
surface. In Sect. IV, the numerical calculations for the IVGDR eigenenergy
and the corresponding EWS sum $m_{1}$ are presented. Conclusions are given
in Sect. V.

\section{Response function and energy-weighted sums}

We consider the response of the nucleus to an external periodic in time
field $U_{\mathrm{ext}}(t)$: 
\begin{equation}
U_{\mathrm{ext}}(t)=\lambda _{0}e^{-i\omega t}\hat{q}+\lambda _{0}^{\ast
}e^{i\omega t}\hat{q}^{\ast },  \label{uext}
\end{equation}%
where $\hat{q}$ is the Hermitian one-particle operator, which depends on
both spatial and isospin coordinates. If we assume  $\lambda _{0}\ll 1$, the
quantum mechanical expectation of the operator $\hat{q}$ is given by the
following form \cite{LaLi5} 
\begin{equation}
\left\langle \hat{q}\right\rangle =\chi (\omega )\lambda _{0}e^{-i\omega
t}+\chi ^{\ast }(\omega )\lambda _{0}^{\ast }e^{i\omega t},  \label{qexp}
\end{equation}%
where $\chi (\omega )$ is the linear response function. We will evaluate the
isovector density-density response function $\chi (\omega )$ by using 
\[
\hat{q}=\sum_{j=1}^{A}\tau _{j}e^{-i\vec{q}_{j}\cdot \vec{r}_{j}}, 
\]%
where $\tau _{j}=1$ for the neutron and $\tau _{j}=-1$ for the proton.

To evaluate the response function $\chi (\omega )$, we use the collisionless
kinetic Landau-Vlasov equation for the isovector excitations 
\begin{equation}
\frac{\partial }{\partial t}\delta f+\vec{v}\cdot \vec{\nabla}_{r}\delta f-%
\vec{\nabla}_{p}f_{\mathrm{eq}}\cdot \vec{\nabla}_{r}(\delta {U}_{\mathrm{%
self}}+{U}_{\mathrm{ext}})=0.  \label{eqlvi}
\end{equation}%
Here, $\delta {f}$ is the isovector variation of the Wigner distribution
function, $\vec{v}$ is the nucleon velocity, $f_{\mathrm{eq}}$ is the
equilibrium distribution function and $\delta {U}_{\mathrm{self}}$ describes
the dynamical component of the selfconsistent mean field.

The solution to Eq. (\ref{eqlvi}) can be written in terms of a plane wave 
\cite{AbKha} 
\begin{equation}
\delta f=-\frac{\partial f_{\mathrm{eq}}}{\partial \varepsilon _{p}}\nu _{%
\vec{q}}(\vec{p})\,e^{i(\vec{q}\cdot \vec{r}-\omega t)},  \label{dfsfind}
\end{equation}%
where $\varepsilon _{p}=p^{2}/2m^{\ast }$, $m^{\ast }$ is the nucleon
effective mass and $\nu _{\vec{q}}(\vec{p})$ is an unknown amplitude. The
variation of the self-consistent field $\delta U_{\mathrm{self}}$ in Eq. (\ref%
{eqlvi}) is obtained through the isovector interaction amplitude $F^{\prime
}(\vec{p},\vec{p}^{\prime })$ as 
\begin{equation}
\delta U_{\mathrm{self}}=\int {\frac{2d{\vec{p}}^{\prime }}{(2\pi \hbar )^{3}%
}\,}F^{\prime }(\vec{p},\vec{p}^{\prime })\,\delta f(\vec{r},{\vec{p}}%
^{\prime };t).  \label{dus}
\end{equation}%
The interaction amplitude $F^{\prime }(\vec{p},\vec{p}^{\prime })$ is
usually parametrized in terms of the Landau constants $F_{k}^{\prime }$ as 
\cite{Mi} 
\begin{equation}
F^{\prime }(\vec{p},\vec{p}^{\prime })=\frac{1}{N_{F}}\sum_{k=0}^{\infty
}\,F_{k}^{\prime }\,P_{k}(\hat{p}\cdot \hat{p}^{\prime }),\,\,\,\,\,\,\,\hat{%
p}=\vec{p}/p,  \label{vsland}
\end{equation}%
where $P_{k}(x)$\ is the Legendre polynomial, $N_{F}$ is the averaged
density of states at the Fermi surface, given by
\begin{equation}
N_{F}=-\int \frac{2d{\vec{p}}}{(2\pi \hbar )^{3}}\ \frac{\partial f_{\mathrm{%
eq}}}{\partial \varepsilon _{p}}\ =\frac{m^{\ast }p_{F}}{\pi ^{2}\hbar ^{3}}
\label{nf}
\end{equation}%
and $p_{F}$ is the Fermi momentum.

The parameter ${F}_{0}^{\prime }$ is related to the isotopic symmetry energy 
$C_{\mathrm{sym}}$ in the Weizs\"{a}cker mass formula \cite{BoMo,Mi} 
\begin{equation}
C_{\mathrm{sym}}=\frac{2}{3}\varepsilon _{F}\,(1+{F}_{0}^{\prime }),
\label{csym}
\end{equation}
with the Fermi energy $\varepsilon _{F}=p_{F}^{2}/2m^{\ast }$. In the following, 
we will assume that 
\begin{equation}
{F}_{0}^{\prime }\neq 0,\quad {F}_{1}^{\prime }\neq 0,\quad F_{l\geq
2}^{\prime }=0.  
\label{fl}
\end{equation}

By using Eq. (\ref{eqlvi}) to get the amplitudes $\nu _{\vec{q}}(\vec{p})$,
we obtain the isovector response function in the form \cite{KoLuKh} 
\begin{equation}
{\chi }(\omega )=\frac{\overline{Q}_{00}(s)}{1-g(s)\ \overline{Q}_{00}(s)}.
\label{resps}
\end{equation}
Here, $s=\omega m^{\ast }/qp_{F}$, 
\begin{equation}
\overline{Q}_{00}(s)=N_{F}Q_{00}(s),\quad Q_{00}(s)=1+\frac{s}{2}\,\ln
\left\vert {\frac{s-1}{s+1}}\right\vert +i\frac{\pi }{2}s\,\theta (1-|s|),
\label{overq00}
\end{equation}%
and%
\begin{equation}
g(s)=-\frac{1}{N_{F}}\left( {F_{0}^{\prime }+\frac{F_{1}^{\prime }}{%
1+F_{1}^{\prime }/3}\ s^{2}}\right) .  \label{kap}
\end{equation}

The frequencies of isovector eigenvibrations [the poles of the response
function (\ref{resps})] can be derived from the dispersion relation 
\begin{equation}
1-g(s)\ \overline{Q}_{00}(s)=0.  \label{dispeq}
\end{equation}
The response function in Eq. (\ref{resps}) allows us to evaluate the
isovector EWS 
\begin{equation}
m_{k}=\frac{1}{\pi }\int_{0}^{\infty }d(\hbar \omega )\ (\hbar \omega )^{k}%
\mathrm{{Im}{\chi }(\omega ).}  \label{mk}
\end{equation}%
Using the dispersion relation between $\mathrm{{Im}{\chi }(\omega )}$ and $%
\mathrm{{Re}{\chi }(\omega )}$, and the asymptotic behavior of $\mathrm{{Re}{%
\chi }(\omega )}$ at $\omega \rightarrow 0$ and $\omega \rightarrow \infty $
limits, one obtains (see also Ref. \cite{LiSt})
\begin{equation}
m_{-1}=\frac{A}{2}\frac{1}{C_{\mathrm{sym}}},\quad m_{1}=\hbar ^{2}\frac{A}{2%
{m}^{\prime }}q^{2},\quad m_{3}=\hbar ^{4}\frac{A}{2}\frac{C_{\mathrm{sym}%
}^{\prime }}{{m^{\prime }}^{2}}q^{4}.  
\label{ews}
\end{equation}%
Here, we introduced the renormalized (because of the Fermi surface distortion
effect) isotopic symmetry energy $C_{\mathrm{sym}}^{\prime }=C_{\mathrm{sym}%
}+8\varepsilon _{F}/15$ and the effective mass $m^{\prime }=m/(1+\kappa
_{NM})$ for the isovector channel, where $\kappa _{NM}$ is the enhancement
factor of the sum rule (for nuclear matter), which is defined by the relation%
\begin{equation}
1+\kappa _{NM}=(1+{F}_{1}^{\prime }/3)/(1+F_{1}/3),  \label{ench1}
\end{equation}%
where $F_{k}$ is the Landau amplitude for the isoscalar channel. In contrast to
the isoscalar excitations, the isovector EWS sum ${m}_{1}$ in Eq. (\ref{ews})
is not model independent in the sense that it depends on the effective
mass ${m}^{\prime }$ and, thereby, on the interaction amplitudes $F_{1}$ and 
${F}_{1}^{\prime }$. We also point out that the EWS ${m}_{k}$ of Eq. (\ref%
{ews}) allows us to evaluate the constrained energy, $E_{\mathrm{constr}}$,
and the scaling energy, $E_{\mathrm{sc}}$, for the IVGDR. Namely, 
\begin{equation}
E_{\mathrm{constr}}=\sqrt{\frac{m_{1}}{m_{-1}}}=\hbar \sqrt{\frac{C_{\mathrm{%
sym}}}{m^{\prime }}}\,q,\quad E_{\mathrm{sc}}=\sqrt{\frac{m_{3}}{m_{1}}}%
=\hbar \sqrt{\frac{{C}_{\mathrm{sym}}^{\prime }}{m^{\prime }}}\,q.
\label{e1e3exp}
\end{equation}

\section{Boundary condition}

For finite nuclei, the dispersion relation [Eq. (\ref{dispeq})] has to be
supplemented by a corresponding boundary condition. The boundary condition
can be viewed as a condition for a balance of the forces acting on the free
nuclear surface%
\begin{equation}
\left. \vec{n}\cdot \vec{F}\right\vert _{S}+\vec{n}\cdot \vec{F}_{S}=0,
\label{bocon}
\end{equation}%
where $\vec{n}$ is the unit vector normal to the nuclear surface $S$, the
internal force $\vec{F}$ is associated with the isovector sound wave and 
$\vec{F}_{S}$ is the isovector surface tension force. The internal force 
$\vec{F}$ is defined by the momentum flux tensor $\delta P_{\alpha\beta}$
inside the nuclear volume. Thus, $F_\alpha=n_\beta\delta P_{\alpha\beta}$
where $\delta P_{\alpha \beta }$ is given by \cite{kosh04,KoKoSh} 
\begin{equation}
\delta P_{\alpha \beta }=\mu _{F}\ \left( \mathbf{\nabla }_{\alpha }\xi
_{\beta }+\mathbf{\nabla }_{\beta }\xi _{\alpha }\right) +\left( C_{\mathrm{%
sym}}\bar{\rho}_{\mathrm{eq}}-\frac{2}{3}\mu _{F}\right) \vec{\nabla}\cdot 
\vec{\xi}\ \delta _{\alpha \beta }.  \label{psound}
\end{equation}%
Here, $\vec{\xi}$ is the displacement field, 
$\bar{\rho}_{\mathrm{eq}}=(\rho_{n,\mathrm{eq}}+\rho_{p,\mathrm{eq}})/2$, and 
\begin{equation}
{\mu }_{F}=\frac{3}{2}\ \rho _{\mathrm{eq}}\varepsilon _{F}\ \frac{s^{2}}{1+{%
F}_{1}^{\prime }/3}\left[ {1-\frac{(1+{F}_{0}^{\prime })(1+{F}_{1}^{\prime
}/3)}{3s^{2}}}\right] .  \label{musf}
\end{equation}

The surface force $\vec{F}_{S}$ in Eq. (\ref{bocon}) is defined as 
$F_{\nu,S}=n_{\nu }\delta P_{S}$, where $\delta P_{S}$ is the pressure caused 
by the isovector polarizations at the nuclear surface \cite{stri83,KoLuKh}, 
given by 
\begin{equation}
\delta P_{S}=\frac{8}{3}\frac{\rho _{\mathrm{eq}}}{r_{0}}Q\ \delta R_{1}.
\label{pshift}
\end{equation}
Here, $r_{0}$ is the mean distance between nucleons, $Q$ is related to the
surface symmetry energy in the Weizs\"{a}cker mass formula \cite{MySw} and $%
\delta R_{1}=R_{0}\alpha _{S}(t)Y_{10}(\hat{r})$. The amplitude $\alpha
_{S}(t)$ of \ the isovector vibrations of the nuclear surface is connected
to the corresponding amplitude $\vec{\xi}$ of the displacement field in a
sound wave.

Finally, from Eqs. (\ref{bocon}), (\ref{psound}) and (\ref{pshift}) we
derive the following secular equation for the wave number $q$ 
\begin{equation}
\left[ \frac{\bar{\rho}_{\mathrm{eq}}}{4}C_{\mathrm{sym}}+\frac{\mu _{F}}{3}-
\frac{\mu _{F}}{x^{2}}\right] \ j_{1}(x)+\left[ \frac{\mu _{F}}{x}-
\frac{2\rho_{\mathrm{eq}}Q}{3qr_{0}(1+\kappa _{NM})}\right] \ j_{1}^{\prime
}(x)=0,  
\label{seceq}
\end{equation}
where $x=qR_{0}$ and $R_{0}=r_{0}A^{1/3}$. In the limit $Q\rightarrow \infty$, 
the boundary condition [Eq. (\ref{seceq})] gives rise to the boundary condition 
$j_{1}^{\prime }(x)=0$ of the Steinwedel-Jensen model \cite{BoMo}. We point
out that the secular equation (\ref{seceq}) for $q$ has to be solved
consistently with the dispersion equation (\ref{dispeq}) for $s$.

\section{Numerical calculations}

Numerical calculations were carried out by using the following set of
nuclear parameters: $r_{0}=1.2$ fm, $F_1=-0.3$ and $F_{0}^{\prime}=1.41$. 
According to Eq. (\ref{csym}), the bulk symmetry energy $C_{\mathrm{sym}}$
is equal to $60$ MeV \cite{BoMo}. The value of the Landau amplitude $%
F_{1}^{\prime }$ will be discussed in the following.

\begin{figure}[tbp]
\vspace{5cm}
\par
\begin{center}
\includegraphics*{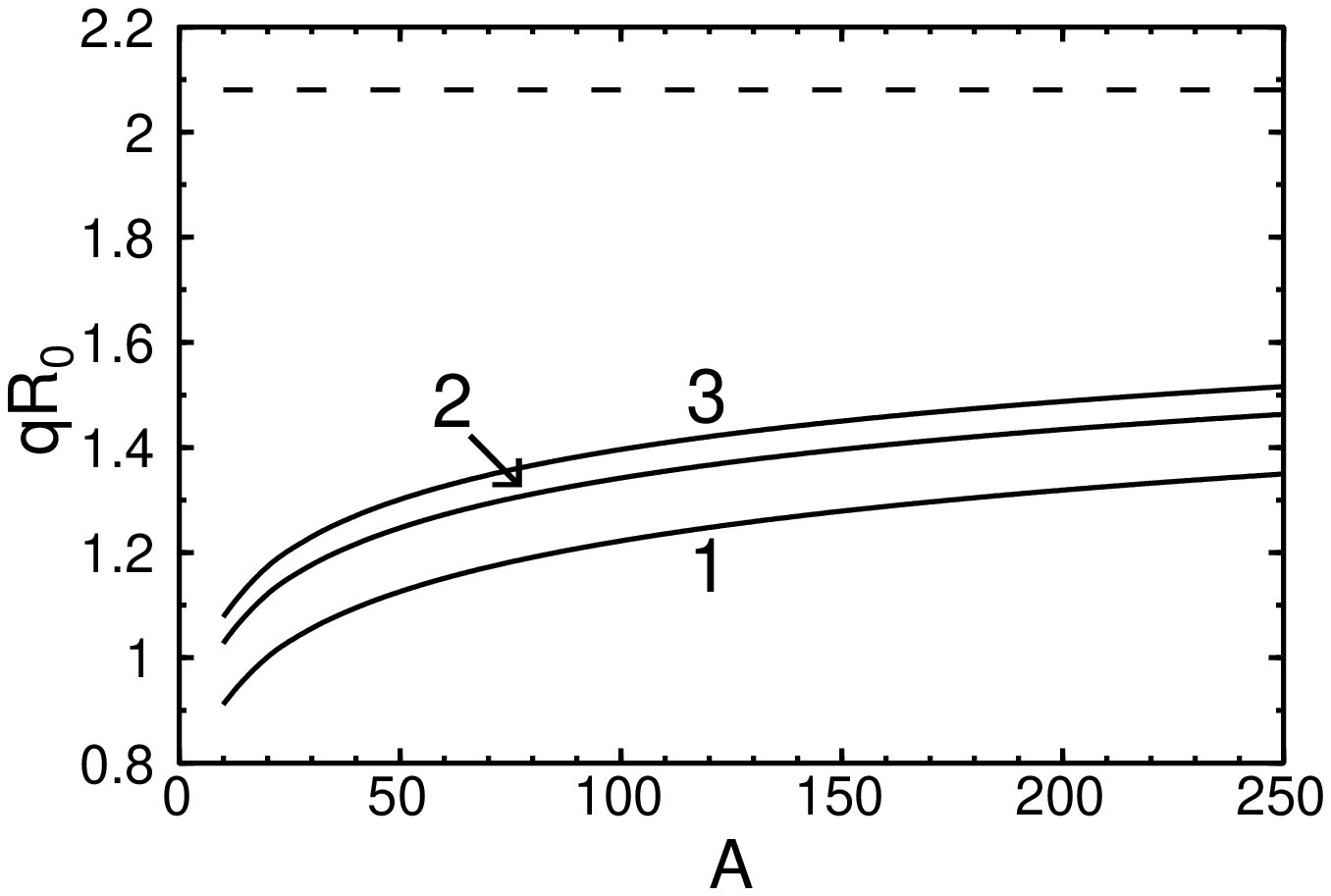}
\end{center}
\caption{Dependence of the value $x=qR_{0}$ on the mass number $A$ obtained
from the secular equation (\protect\ref{seceq}) for $Q=10.5$ MeV and $%
F_{1}^{\prime }=1.1$.}
\label{fig1}
\end{figure}

In \figurename\ \ref{fig1}, the solid curve 1 shows the $A$ dependence of the
value $x$ obtained from the secular equation (\ref{seceq}); it is
consistent with the dispersion relation [Eq. (\ref{dispeq})] including all
multipolarities $l$ of the Fermi surface distortions. Curve 3 represents
the analogous result but for the velocity-independent nuclear forces (i.e.,
for $F_{1}=0$ and $F_{1}^{\prime }=0$). Curve 2 shows a solution to the
secular equation (\ref{seceq}), when one takes into account the Fermi
surface distortions up to quadrupole order (scaling approximation with $%
l\leq 2$) \cite{stri83}. In the last case, instead of solving the dispersion
equation (\ref{dispeq}), we have used the expression $s^{2}=(9/5+F_{0}^{%
\prime })/3$, which gives the dimensionless sound velocity $s$ for $l\leq 2$ 
\cite{KoKoSh}. As seen from \figurename\ \ref{fig1}, the value of $x$ is
significantly smaller than the corresponding one obtained within the
Steinwedel-Jensen model (dashed line in \figurename\ \ref{fig1}).

\begin{figure}[tbp]
\vspace{5cm}
\par
\begin{center}
\includegraphics*{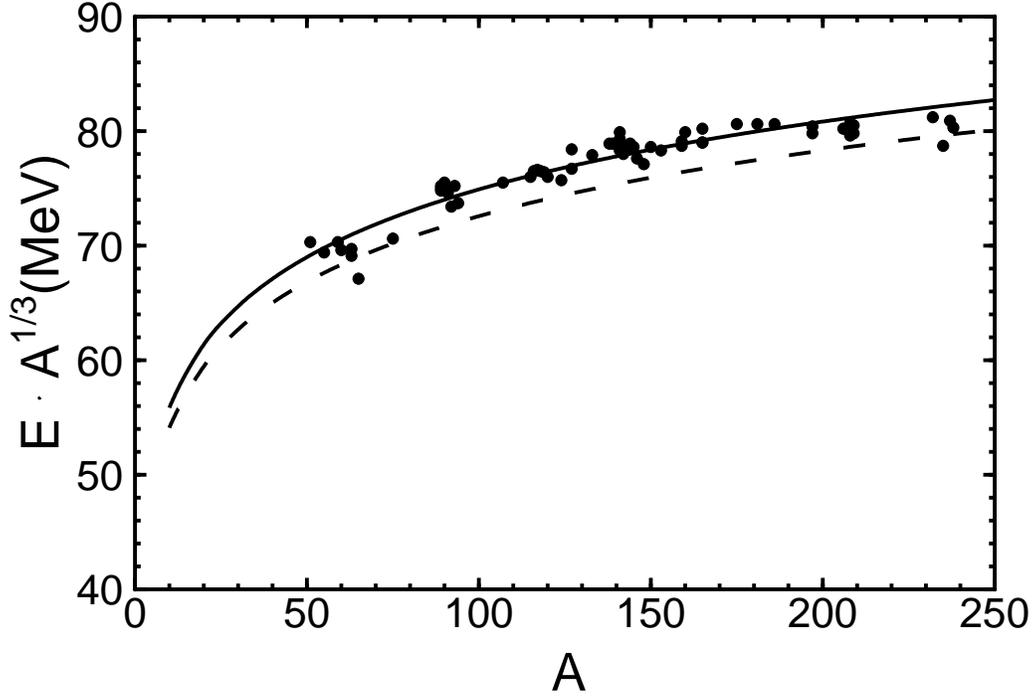}
\end{center}
\caption{Dependence of the IVGDR energy on the mass number $A$. The dashed
line is the calculation that includes Fermi surface deformations up to 
quadrupole order [scaling approximation: see $E_{\mathrm{sc}}$ in Eq. 
(\protect\ref{e1e3exp})]; the solid line was obtained from the dispersion
equation (\protect\ref{dispeq}) and the secular equation (\protect\ref{seceq}%
). The dots are the experimental data taken from Ref. \protect\cite{BeFu}.}
\label{fig2}
\end{figure}

In \figurename\ \ref{fig2} we show a dependence of the IVGDR energy
(multiplied by the factor $A^{1/3}$) on the mass number $A$. The calculations
have been performed at $Q=10.5$ \textrm{MeV} and $F_{1}^{\prime }=1.1$. The
solid line is the eigenenergy obtained from the dispersion equation (\ref{dispeq}) 
supplemented by the boundary condition [Eq. (\ref{seceq})]. The dashed
line in \figurename\ \ref{fig2} was obtained from the EWS definition of the
scaling energy, $E_{\mathrm{sc}}$, of Eq. (\ref{e1e3exp}). A significant
upward shift of the exact eigenenergy (solid line) with respect to the
scaling one, $E_{\mathrm{sc}}$, is due to the Fermi surface distortions of
the higher multipolarities ($l>2$) contributed to the dispersion equation (\ref%
{dispeq}). The values of the Landau amplitude $F_{1}^{\prime }=1.1$ and the
parameter $Q=10.5$ \textrm{MeV} have been chosen such that the best fit of
the $A$ dependence of the eigenenergies to the experimental data is obtained.

Let us now consider the enhancement factor $\kappa $ of the isovector EWS $%
m_{1}$. For infinite nuclear matter, it is given by the value of $\kappa
_{NM}$ in Eq. (\ref{ench1}). The experimental determination of the
enhancement factor $\kappa $ is connected to the investigation of the
photoabsorption cross section $\sigma _{\mathrm{abs}}(\omega)$ of $\gamma$ 
quanta. For the velocity-independent forces, the isovector EWS $m_{1}$ is
model independent and reads (TRK sum rule) \cite{RiSh} 
\begin{equation}
\widetilde{m}_{1,TRK}=\int\limits_{0}^{\infty}d(\hbar\omega)\ \sigma_{\mathrm{abs}}
(\omega)=\frac{2\pi^{2}\hbar\ e^{2}}{mc}\frac{NZ}{A}.
\label{trkr}
\end{equation}
The photoabsorption cross section $\sigma_{\mathrm{abs}}(\omega )$ can be
expressed in terms of the strength function $S(\omega)=\mathrm{{Im}{\chi}(\omega)/\pi}$ 
as follows \cite{diko99}: 
\begin{equation}
\sigma _{\mathrm{abs}}(\omega )={\frac{4\pi ^{2}e^{2}}{cq_{0}^{2}(A)\rho _{0}%
}}{\frac{NZ}{A}}\omega \ S(\omega),  
\label{sigabs}
\end{equation}%
where the wave number $q_{0}(A)$ has to be found from Eq. (\ref{seceq}) in
the limit of velocity-independent forces, i.e., at $F_{1}=0$ and $%
F_{1}^{\prime }=0$.

In a general case of velocity-dependent nuclear forces, by using Eq. 
(\ref{ews}) for $m_{1}$ and Eqs. (\ref{trkr}) and (\ref{sigabs}), we
generalize the TRK sum rule in the form 
\begin{equation}
\widetilde{m}_{1}(A)=\frac{2\pi ^{2}\hbar e^{2}}{mc}\frac{NZ}{A}\left[ \frac{%
q_{1}(A)}{q_{0}(A)}\right] ^{2}(1+\kappa _{NM}),  \label{trknew}
\end{equation}%
where $q_{1}(A)$ is derived by Eq. (\ref{seceq}) at $F_{1}\neq 0$ and $%
F_{1}^{\prime }\neq 0$.

\begin{figure}[tbp]
\vspace{5cm}
\par
\begin{center}
\includegraphics*{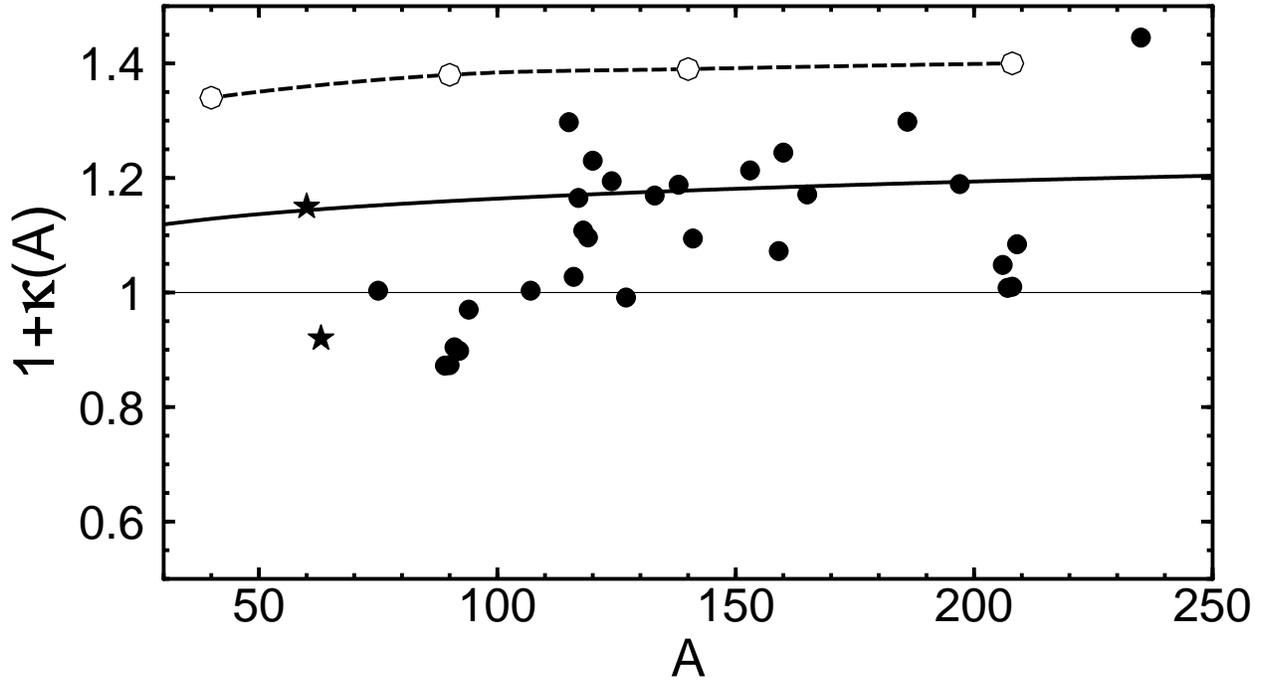}
\end{center}
\caption{Dependence of the enhancement factor $1+\protect\kappa (A)$ for
IVGDR on the mass number $A$. The calculation result was obtained for a 
Landau amplitude $F_{1}=-0.3,\ F_{1}^{\prime }=1.1$ (solid curve). The
dashed line is the microscopic RPA calculations with Skyrme forces SkM$%
^{\ast }$ from Refs. \protect\cite{LiSt,befl75}. The solid points are the
experimental data of the Livermore group from Ref. \protect\cite{BeFu}. Two
points (noted by the symbol $\star $) were obtained by the inclusion of the
contribution from the $(\protect\gamma ,p)$\ cross section.}
\label{fig3}
\end{figure}

In \figurename\ \ref{fig3}, the EWS enhancement factor $1+\kappa (A)$ is
plotted as a function of the mass number $A$: 
\begin{equation}
\frac{\widetilde{m}_{1}(A)}{\widetilde{m}_{1,TRK}}
=\left[{\frac{q_{1}(A)}{q_{0}(A)}}\right]^{2}(1+\kappa_{NM})
=1+\kappa (A).  
\label{ka}
\end{equation}
The exceedance of the $100{\%}$ sum rule for $\widetilde{m}_{1}(A)$, which is
observed for the IVGDR, is caused by the dependence of the effective
nucleon-nucleon interaction on the nucleon velocity. For the value of the
isovector amplitude $F_{1}^{\prime }=1.1$, one can fit (on average) the
results of our calculations of $1+\kappa (A)$ (solid curve in \figurename\ 
\ref{fig3}) to the experimental data of the Livermore group \cite{BeFu}. Our
estimate to the enhancement factor $\kappa (A)$ is about $10\%$ for light
nuclei and increases to $20\%$ for heavy nuclei. This result is consistent
with the general conclusion of Ref. \cite{BeFu}; that is, for $A>100$, the
experimental data center is at about $1.2$ TRK sum-rule units. For the mass region 
$A<70$, the TRK sum rule is not exhausted. As noted in Ref. \cite{BeFu},
this no doubt results from the neglect of the $(\gamma ,p)$ channel
contribution for these nuclei. It was reported recently that inclusion of
the contribution from the $(\gamma ,p)$ cross section increases the EWS
exhaustion from $0.87$ to $1.15$\ for $^{60}$Ni and from $0.64$\ to $0.92$\
for $^{63}$Cu \cite{Be}. The corresponding new data are shown in \figurename%
\ \ref{fig3} by the symbol $\star$. The result of a microscopic RPA calculation
of the enhancement factor is shown in \figurename\ \ref{fig3} as a dashed
line (see Refs. \cite{LiSt,befl75}). We point out that both our Fermi-liquid
approach and the RPA prediction give very similar $A$ dependence of the
enhancement factor $1+\kappa (A)$. However, it is seen from \figurename\ 
\ref{fig3} that the RPA calculation overestimates the magnitude of $\kappa (A)$
(see also Refs. \cite{tror87,krtr80,sabo04}). The RPA result for $\kappa (A)$
can be improved by a fit of the relevant parameters $t_{1}$ and $t_{2}$ in
the Skyrme forces.

The corresponding study has been recently performed within the microscopic
Hartree-Fock plus RPA approach in Ref. \cite{trco08}, where a best value
for $\kappa (A)$ of $0.22\pm 0.04$ was deduced from a fit to the experimental
data for the nucleus $^{208}$Pb. However, reducing the enhancement factor
by the variation of the Skyrme force parameters, one can expect a significant 
change of the $A$ behavior (slope) of the corresponding curve $\kappa (A)$ 
(dashed line in \figurename\ \ref{fig3}). Unfortunately, in the nuclear 
literature, this kind of analysis has not yet been carried out. We also
note that the latest microscopic RPA calculations presented in Refs. 
\cite{nest08,trco08} are related to nuclear matter and the nucleus $^{208}$Pb
only. They do not show an $A$ dependence of the enhancement factor, but 
they do give a demonstration of significant variation of the dependence of the
enhancement factor on the choice of the Skyrme force parametrization (see
the last column in Table I of Ref. \cite{trco08}).

\section{Conclusions}

In conclusion, we wish to comment that it was conceptually important for us
to achieve a description of the $A$ dependence of both the IVGDR energy and
the enhancement factor simultaneously (see FIGs. 2 and 3). In our approach,
we used appropriate boundary conditions that allowed us to
combine both the Steinwedel-Jensen and Goldhaber-Teller models. A similar
problem was considered earlier in Ref. \cite{stri83} but within the scaling
approximation only. Thus, an inclusion of the effective isovector surface
stiffness $Q$ into the boundary condition [Eq. (\ref{seceq})] leads to the $A$ 
dependence of the value $qR_{0}$, which becomes significantly smaller than
the Steinwedel-Jensen's estimate $qR_{0}=2.08$ \cite{BoMo}. Using the
obtained value of $qR_{0}$, we have described the IVGDR energies within the
Landau kinetic theory quite well. Fitting the slope of the energy dependence
on the mass number $A$ to the experimental data, we have estimated the value
of the effective isovector surface stiffness as $Q\simeq 11$ MeV.

The dependence of the effective nucleon-nucleon interaction on the nucleon
velocity causes the $100{\%}$ exhaustion of the TRK sum rule to be exceeded 
for the IVGDR. Within Landau kinetic theory, the EWS enhancement factor 
$\kappa_{NM}$ in infinite nuclear matter depends on the interaction amplitudes 
$F_{1}$ and $F_{1}^{\prime }$. In finite nuclei, the $A$ dependence of the
EWS enhancement factor $\kappa (A)$ occurs because of the boundary condition 
[Eq. (\ref{seceq})]. The value of $\kappa (A)$ increases with $A$. A fit of the
enhancement factor to the proper experimental data leads to a value for the
isovector Landau amplitude of $F_{1}^{\prime }\simeq 1.1$. The obtained value
of $F_{1}^{\prime }$ exceeds the estimate $F_{1}^{\prime}=0.5-0.7$ derived 
earlier from Skyrme forces for infinite nuclear matter \cite{krtr80,lilu91,Mi}. 
This exceedance appears since our derivation of $F_{1}^{\prime }$ is related 
to the interior of the finite nucleus. For finite nuclei, the Landau amplitudes 
$F_{k}$ and $F_{k}^{\prime }$ are $r$-dependent ones with a bump within the 
nuclear surface \cite{lilu91,Mi}. This effectively increases the bulk values 
of $F_{k}$ and $F_{k}^{\prime}$ in the limit of the sharp nuclear surface 
assumed in this paper.

We show that the value of the enhancement factor is about $10\%$ for light
nuclei and reaches approximately $20\%$ for heavy nuclei. For nuclei with 
$A>100$, our results are close to the experimental data from Livermore
discussed in Ref. \cite{BeFu} and they are in agreement with the best value
for the enhancement factor obtained for the nucleus $^{208}$Pb in Ref. 
\cite{trco08}. We note also that the $A$ behavior of the enhancement factor 
obtained in our Fermi-liquid approach is similar to the one derived from the 
microscopic RPA calculations (see dashed line in FIG. 3), but, as was previously
mentioned, the RPA results show a strong variation of the enhancement
factor with the choice of the Skyrme force parametrization \cite{trco08}.

\newpage

\newpage

\newpage

\end{document}